\documentclass[11pt]{article}
\usepackage{graphicx}
\usepackage{epsfig}

\newcommand{\BABARPubYear}    {04}

\newcommand{\BABARProcNumber} {112}
\newcommand{\SLACPubNumber} {10781}

\RequirePackage{xspace}
\usepackage{relsize}
\def\babar{\mbox{\slshape B\kern-0.1em{\smaller A}\kern-0.1em
    B\kern-0.1em{\smaller A\kern-0.2em R}}}
\def\ccbar {\ensuremath{c\overline c}\xspace}
\def\pep2{PEP-II}
\def\Dz      {\ensuremath{D^0}\xspace}
\def\Ds      {\ensuremath{D^+_s}\xspace}
\def\piz   {\ensuremath{\pi^0}\xspace}
\def\pip   {\ensuremath{\pi^+}\xspace}
\def\pim   {\ensuremath{\pi^-}\xspace}
\def\invfb   {\ensuremath{\mbox{\,fb}^{-1}}\xspace}
\newcommand{\jprd}      [1]  {\jprBase\ D~{\bf #1}}
\newcommand{\jprBase}        {Phys.\ Rev.\xspace}
\newcommand{\jprl}      [1]  {\jprlBase\ {\bf #1}}
\newcommand{\jprlBase}       {Phys.\ Rev.\ Lett.\xspace}
\def\DsTT{D_{sJ}^*(2317)^+}

\def\DsFE{D_{sJ}(2460)^+}

\long\def\inst#1{\par\nobreak\kern 4pt\nobreak
    {\it #1}\par\vskip 10pt plus 3pt minus 3pt}

\begin{document}
{\pagestyle{empty}

\begin{flushright}
SLAC-PUB-\SLACPubNumber \\
\babar-PROC-\BABARPubYear/\BABARProcNumber \\
October, 2004 \\
\end{flushright}

\par\vskip 4cm

\begin{center}
\Large \bf Charm decays at \babar
\end{center}
\bigskip

\begin{center}
\large 
M. Charles\\
University of Iowa\\
SLAC M/S 35, 2575 Sand Hill Road, Menlo Park, CA 94025\\
(representing the \babar\ Collaboration)
\end{center}
\bigskip \bigskip

\begin{center}
\large \bf Abstract
\end{center}
    The results of several studies of charmed mesons and baryons at \babar \ are
    presented.
    First, searches for the rare decays $D^0 \rightarrow l^+ l^-$ are presented
    and new upper limits on these processes are established.
    Second, a measurement of the branching fraction of the isospin-violating hadronic decay
    $D_s^*(2112)^+ \rightarrow D_s^+ \pi^0$ relative to the radiative decay
    $D_s^*(2112)^+ \rightarrow D_s^+ \gamma$ is made.
    Third, the decays of $D_{sJ}^*(2317)^+$ and $D_{sJ}(2460)^+$
    mesons are studied and ratios of branching fractions are measured.
    Fourth, Cabibbo-suppressed decays of the $\Lambda_c^+$ are examined
    and their branching fractions measured relative to Cabibbo-allowed modes.
    Fifth, the $\Xi_c^0$ is studied through its decays to
    $\Xi^- \pi^+$ and $\Omega^- K^+$; in addition to measuring the ratio of
    branching fractions for $\Xi_c^0$ produced from the \ccbar continuum, the
    uncorrected momentum spectrum is measured, providing clear confirmation of
    $\Xi_c^0$ production in $B$ decays.

\vfill
\begin{center}
Contributed to the Proceedings of the 32$^{nd}$ International 
Conference on High Energy Physics, \\
8/16/2004---8/22/2004, Beijing, China
\end{center}

\vspace{1.0cm}
\begin{center}
{\em Stanford Linear Accelerator Center, Stanford University, 
Stanford, CA 94309} \\ \vspace{0.1cm}\hrule\vspace{0.1cm}
Work supported in part by Department of Energy contract DE-AC02-76SF00515.
\end{center}

\section{Introduction}
\label{sec:intro}

The data for these analyses are collected with the \babar \ detector\cite{ref:babar}
at the \pep2\ asymmetric $e^+e^-$ collider, operating at the $\Upsilon(4S)$
resonance and at a center-of-mass (CM) energy $\sim 40$~MeV below it.
At these energies there is copious production of \ccbar pairs from the
continuum\cite{ref:physbook}. Combined with high integrated luminosity,
this makes \babar \ an excellent laboratory for studying charm production and decays.
Detailed discussions of the topics in this paper may be found in the conference
submissions\cite{bib:Dtoll:eprint,bib:Dsstar:eprint,bib:DsJ:eprint,bib:Lambdac:eprint,bib:Xic:eprint}.
All results are preliminary.

\section{Search for $D^0 \rightarrow l^+ l^-$}

\begin{table}[t]
  \caption{
    The expected background,
    the number of observed events,
    the corresponding branching fraction upper limit at the 90\% confidence level (preliminary),
    and the previous upper limit for each decay mode.
  }
  \label{tab:Dtoll:results}
  \begin{center}
  \begin{tabular}{|l|ccc|} \hline
      & $\Dz\to e^+e^-$
      & $\Dz\to \mu^+\mu^-$
      & $\Dz\to e^\pm\mu^\mp$
      \\\hline 
      Expected $D^0 \rightarrow h^+ h^-$ backgrnd. ($h = \pi,K$) & $0.02$ & $3.3\pm0.3$ & $0.21$ \\
      Expected combinatoric backgrnd. & $2.2\pm 0.4$ & $1.3\pm0.3$ & $1.9\pm0.4$ \\
      Candidates observed in signal window & 3 & 1 & 0 \\
      90\% confidence level upper limit ($10^{-6}$) & 1.2 & 1.3 & 0.81 \\
      Previous upper limit ($10^{-6}$) & 6.2 & 2.5 & 8.1 \\ \hline
  \end{tabular}
  \end{center}
\end{table}

\begin{table}[t]
  \caption{Preliminary ratios of branching fractions in the $D_{sJ}$ system.}
  \label{tab:DsJ:results}
  \begin{center}
  \begin{tabular}{|ll|}
    \hline
    ${\mathcal{B}(\DsTT\to\Ds\gamma)}/{\mathcal{B}(\DsTT\to\Ds\piz)}$ & $< 0.17$ @ 95\% C.L. \\
    ${\mathcal{B}(\DsTT\to\Ds\pip\pim)}/{\mathcal{B}(\DsTT\to\Ds\piz)}$ & $< 0.11$  @ 95\% C.L. \\
    \hline
    ${\mathcal{B}(\DsFE\to\Ds\gamma)}/{\mathcal{B}(\DsFE\to\Ds\piz\gamma)}$ & $0.375 \pm 0.054 \pm 0.057$ \\
    ${\mathcal{B}(\DsFE\to\Ds\pi^+\pi^-)}/{\mathcal{B}(\DsFE\to\Ds\piz\gamma)}$ & $0.082 \pm 0.018 \pm 0.011$ \\
    ${\mathcal{B}(\DsFE\to\Ds\piz)}/{\mathcal{B}(\DsFE\to\Ds\piz\gamma)}$ & $< 0.002$ @ 95\% C.L. \\
    ${\mathcal{B}(\DsFE\to\DsTT\gamma)}/{\mathcal{B}(\DsFE\to\Ds\piz\gamma)}$ & $< 0.23$ @ 95\% C.L. \\
    \hline
  \end{tabular}
  \end{center}
\end{table}

In the Standard Model (SM), the decay modes\footnote{
  Unless otherwise stated, the use of charge conjugate modes is implied throughout.
} $D^0 \rightarrow e^+ e^-$ and $D^0 \rightarrow \mu^+ \mu^-$ are strongly
suppressed by the GIM mechanism\cite{ref:Dtoll:GIM}; their branching fractions
are estimated to be $10^{-23}$ and $3 \times 10^{-13}$
respectively\cite{ref:Dtoll:burdman}. The lepton
flavour violating decay $D^0 \rightarrow e^{\pm} \mu^{\mp}$ is strictly
forbidden in the SM. However, new physics could enhance
these rates\cite{ref:Dtoll:burdman}.
The results of searches for all three modes using 122~\invfb of data are presented in this paper.

The event selection comprises invariant mass cuts and particle identification (PID) criteria.
In order to improve the purity of the $D^0$ candidate sample, it is required
that an additional $\pi^+$ track be present and that the $D^0$ candidate
be consistent with a $D^{*+} \rightarrow D^0 \pi^+$ decay. The expected backgrounds
and numbers of observed events are shown in Table~\ref{tab:Dtoll:results},
along with the preliminary upper limits\cite{ref:Dtoll:extended_feldman_cousins}.
The limits are normalized to the kinematically similar $D^0 \rightarrow \pi^+ \pi^-$ mode\cite{ref:pdg2002}.
These represent significant improvements on the previous limits\cite{ref:Dtoll:e791,ref:Dtoll:herab},
and further restrict possible new physics contributions to these decay processes.

\section{Measurement of
  $\mathcal{B}(D_s^*(2112) \rightarrow D_s^+ \pi^0) /
   \mathcal{B}(D_s^*(2112) \rightarrow D_s^+ \gamma)$
}

Compared to $D_s^*(2112)^+ \rightarrow D_s^+ \gamma$, the hadronic
decay $D_s^*(2112)^+ \rightarrow D_s^+ \pi^0$ is expected to be
isospin-suppressed. Using chiral perturbation theory and
assuming virtual $\eta$ emission, sizeable hadronic decay rates have
been predicted\cite{ref:Dsstar:cho_and_wise}, and were subsequently
measured by the CLEO collaboration\cite{ref:Dsstar:cleo}.
In this section, a \babar \ measurement of the ratio using
90.4~\invfb of data is presented.

The $D_s^+$ mesons are reconstructed using the decay $D_s^+ \rightarrow \phi \pi^+$,
$\phi \rightarrow K^+ K^-$. PID requirements are imposed upon the kaons.
The $D_s^+$ candidate is then combined with a $\gamma$ or $\pi^0$ which
satisfies selection criteria including mass, energy, and (for $\pi^0$) helicity angle requirements
to form a $D_s^*(2112)^+$ candidate. The mass offset spectra are shown in
Figure~\ref{fig:Dsstar:mass}. After correcting for efficiency and taking into
account systematic uncertainties (predominantly due to momentum dependence of the
efficiencies and limited statistics of the Monte Carlo simulation),
the preliminary ratio of branching fractions is measured to be\footnote{
  When a value is quoted with two uncertainties in this paper, the first is
  statistical and the second is systematic. When an upper limit
  is quoted, the limit incorporates the effects of both the statistical
  and the systematic uncertainty.
} $0.0621 \pm 0.0049 \pm 0.0063$.
As a crosscheck, the non-isospin-suppressed decay $D^{*0} \rightarrow D^0 \pi^0$
relative to $D^{*0} \rightarrow D^0 \gamma$ was also measured. This was found to be
$1.740 \pm 0.020 \pm 0.125$, consistent with the existing world average\cite{ref:pdg2004}.

\begin{figure}
  \begin{center}
  \epsfig{file=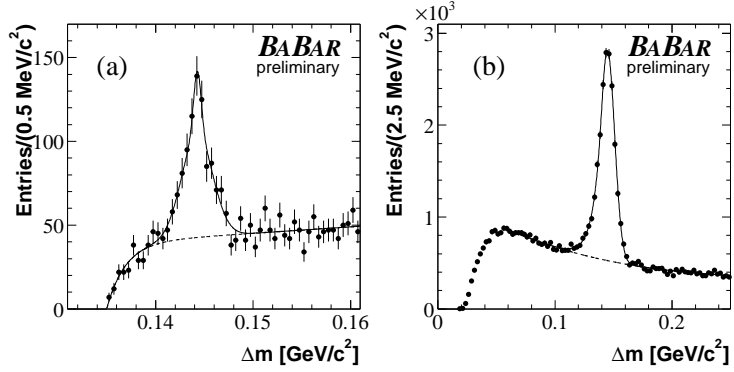, width=0.6\textwidth}
  \end{center}
  \caption{$D_s^{*}(2112)^+$ signals:
    (a)~$m(K^+K^-\pi^+\pi^0)-m(K^+K^-\pi^+)$;
    (b)~$m(K^+K^-\pi^+\gamma)-m(K^+K^-\pi^+)$.
    The dots represent data points. The solid line shows the fitted function, and
    the dashed line indicates estimated background contribution.}
  \label{fig:Dsstar:mass}
\end{figure}

\section{$D_{sJ}^+$ decays}

The unexpected discovery of the $D_{sJ}^*(2317)^+$ and $D_{sJ}(2460)^+$
mesons\cite{Aubert:2003fg,Besson:2003cp} and subsequent studies
of their decays\cite{Abe:2003jk,Krokovny:2003zq,Aubert:2003pe}
have re-awoken interest in charm meson spectroscopy. In this section,
new \babar \ measurements of ratios of their branching fractions are
presented. These measurements are made with inclusive $D_{sJ}$
samples in 125~\invfb of data. In each case, $D_{sJ}$ candidates
are required to have a CM momentum ($p^*$) of at least
3.2~GeV/$c$. This suppresses $D_{sJ}$ production from $B$ decays.
The preliminary ratios of branching fractions and upper limits obtained
are shown in Table~\ref{tab:DsJ:results}.
Details of the fitting procedure, including proper handling of
reflections, are given in the conference submission\cite{bib:DsJ:eprint}.
In addition, the $D_s^+ \pi^{\pm}$ invariant mass spectra
were examined: no evidence of any narrow structure is found close
to the $D_{sJ}^*(2317)$ mass.
This is consistent with the $D_{sJ}^*(2317)$ being an isosinglet state.

\begin{table}
  \caption{Preliminary $\Lambda_c^+$ branching fraction ratios.}
  \label{tab:Lambdac:results}
  \begin{center}
  \begin{tabular}{|ll|}
    \hline
    $\frac{\mathcal{B}(\Lambda_c^+ \rightarrow \Lambda K^+)}
      {\mathcal{B}(\Lambda_c^+ \rightarrow \Lambda \pi^+)}$ 
      &  $0.044 \pm 0.004 \pm 0.002$ \\
    $\frac{\mathcal{B}(\Lambda_c^+ \rightarrow \Sigma^0 K^+)}
      {\mathcal{B}(\Lambda_c^+ \rightarrow \Sigma^0 \pi^+)}$
      & $0.040 \pm 0.005 \pm 0.004$ \\
    $\frac{\mathcal{B}(\Lambda_c^+ \rightarrow \Lambda K^+ \pi^+ \pi^-)}
      {\mathcal{B}(\Lambda_c^+ \rightarrow \Lambda \pi^+)}$
      & $0.266 \pm 0.027 \pm 0.032$ \\
    $\frac{\mathcal{B}(\Lambda_c^+ \rightarrow \Sigma^0 K^+ \pi^+ \pi^-)}
      {\mathcal{B}(\Lambda_c^+ \rightarrow \Sigma^0 \pi^+)}$
      & $ < 0.039$ @ 90\% C.L.\\
      \hline
  \end{tabular}
  \end{center}
\end{table}

\begin{figure}
  \begin{center}
  \epsfig{file=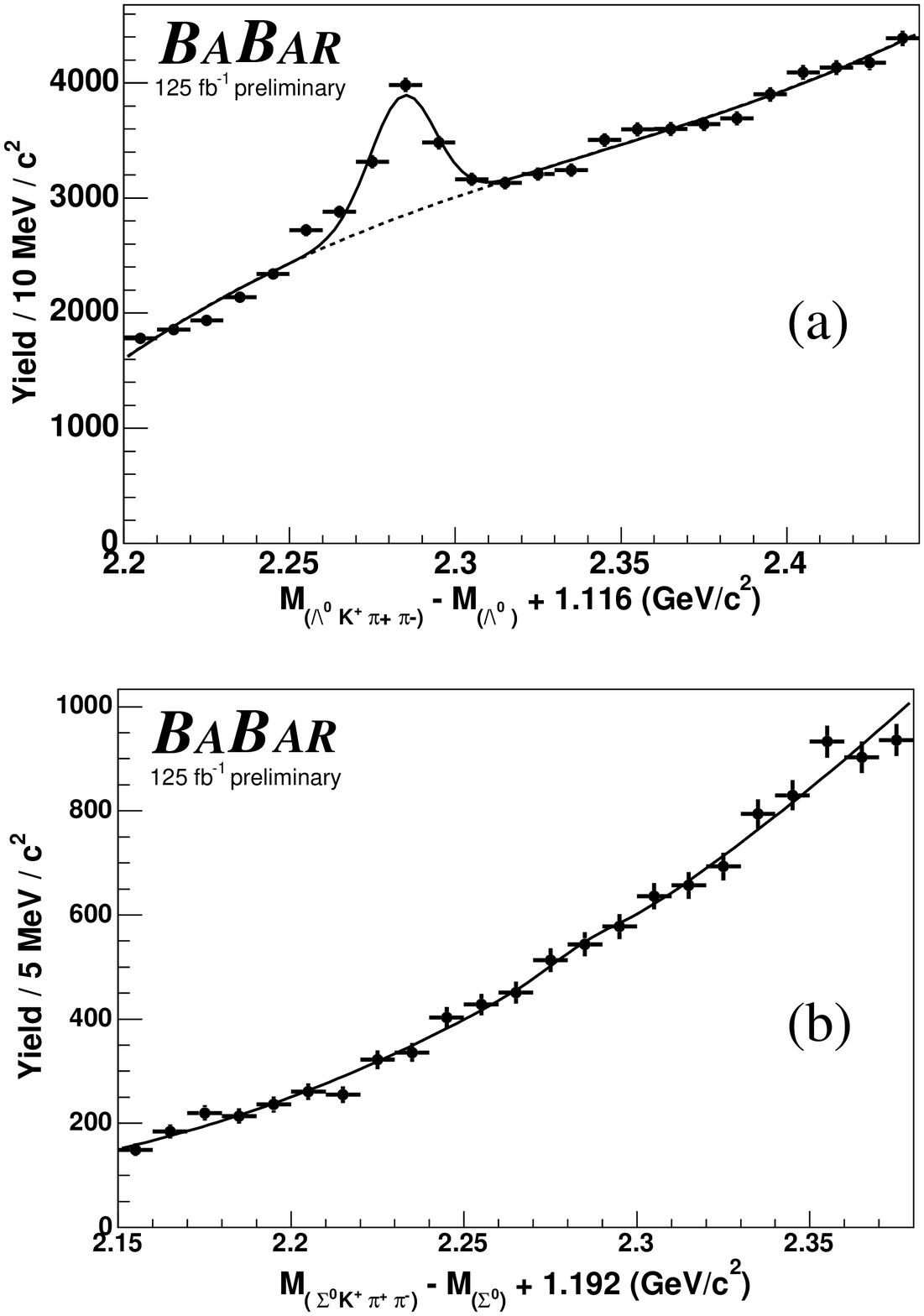, width=0.6\textwidth}
  \end{center}
  \caption{The invariant mass spectra for (a) $\Lambda K^+ \pi^+ \pi^-$,
    (b) $\Sigma^0 K^+ \pi^+ \pi^-$. A clear $\Lambda_c^+$ peak can be seen in (a),
     but not in (b).}
  \label{fig:Lambdac:firstobs}
\end{figure}

\begin{figure}
  \begin{center}
  \epsfig{file=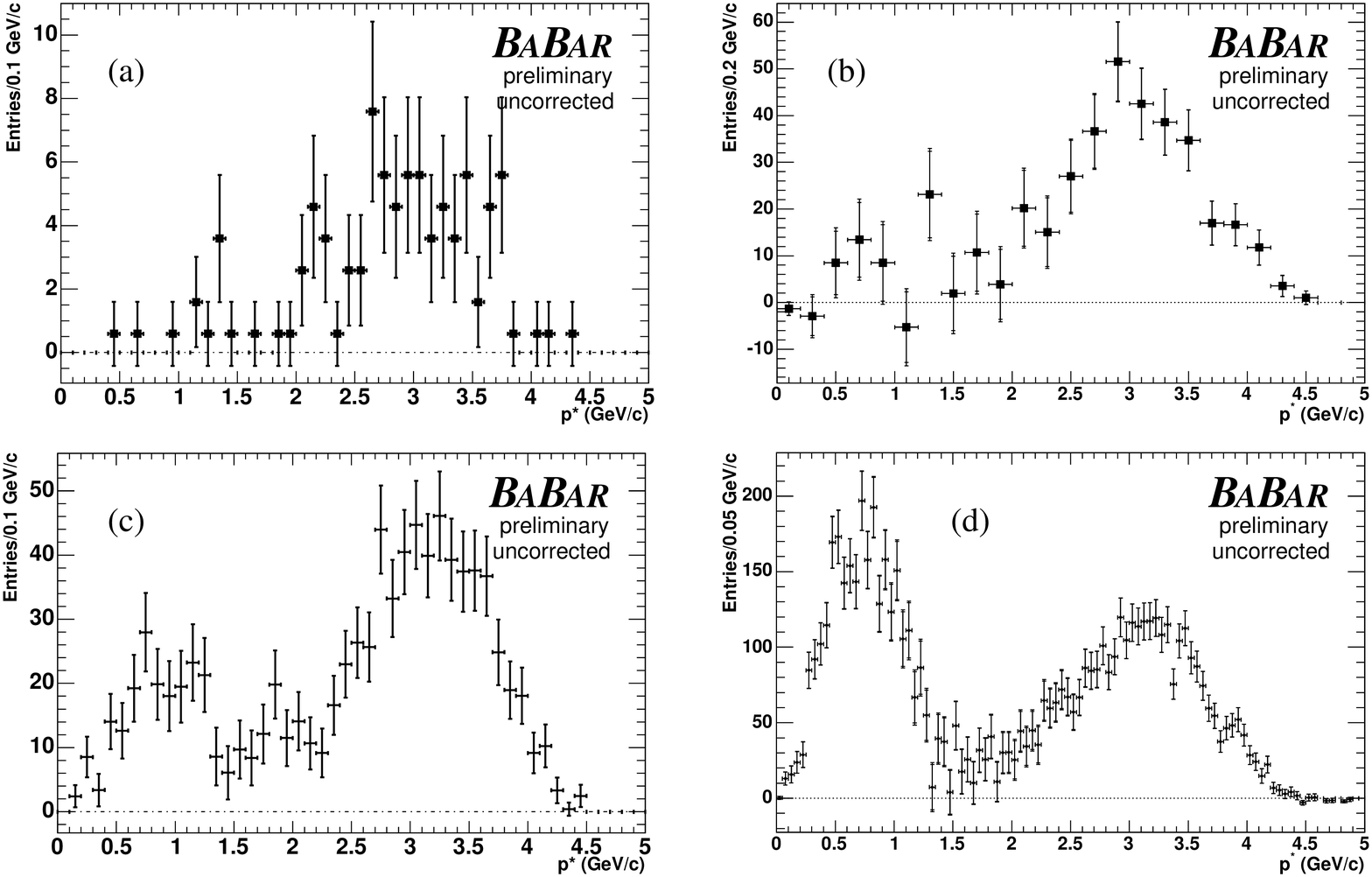, width=0.6\textwidth}
  \end{center}
  \caption{Uncorrected $p^*$ distributions. 
    The $\Xi_c^0 \rightarrow \Omega^- K^+$ spectra are shown in (a) and (c),
    and the $\Xi_c^0 \rightarrow \Xi^- \pi^+$ spectra are shown in (b) and (d).
    Figures (a) and (b) correspond to 11~\invfb of data taken at CM energy
    $\sim 40$~MeV below the $\Upsilon(4S)$, while (c) and (d) correspond to
    104~\invfb of data taken at the $\Upsilon(4S)$ resonance.
    }
  \label{fig:Xic:pdist}
\end{figure}

\section{Cabibbo-suppressed $\Lambda_c^+$ decays}

In this section, four new measurements of Cabibbo-suppressed $\Lambda_c^+$ decays
using 125~\invfb of data are presented.
The decay products include a $\Lambda (p \pi^-)$ or $\Sigma^0 (\Lambda \gamma)$
plus one or more charged pions or kaons. Background is rejected through
mass and flight distance cuts on the hyperons, PID requirements, an
energy threshold for $\gamma$ candidates, and $p^*$ thresholds for $\Lambda_c^+$ candidates.
In each case, the branching fraction is measured relative to a similar,
Cabibbo-allowed mode. The preliminary results, corrected for
efficiency, are given in Table~\ref{tab:Lambdac:results}.
These represent significant improvements over existing results\cite{bib:Lambdac:Belle}.
This is the first observation of the decay
$\Lambda_c^+ \rightarrow \Lambda K^+ \pi^+ \pi^-$; the mass spectrum for
this mode is shown in Figure~\ref{fig:Lambdac:firstobs}~(a). 
The decay mode $\Lambda_c^+ \rightarrow \Sigma^0 K^+ \pi^+ \pi^-$
appears to be strongly suppressed relative to $\Lambda_c^+ \rightarrow \Sigma^0 K^+$
(Figure~\ref{fig:Lambdac:firstobs}); at present there is no explanation
of this intriguing feature.

\section{$\Xi_c^0$ production and decays}

In this section, the ratio of $\Xi_c^0$ branching fractions to
the $\Omega^- K^+$ and $\Xi^- \pi^+$ final states is measured using 116~\invfb of data.
In addition, the production of
$\Xi_c^0$ in $B$ decays is observed. Although copious production of
$\Xi_c^0$ and $\Xi_c^+$ in $B$ decays has been
predicted\cite{bib:Xic:theory_BtoXic_production}, this process has
been observed previously only by CLEO, with a significance of
$\sim 3 \sigma$ in the $\Xi_c^0 \rightarrow \Xi^- \pi^+$ decay mode and
$\sim 4 \sigma$ in a related $\Xi_c^+$ decay mode\cite{bib:Xic:cleo_BtoXic}.

$\Xi_c^0$ candidates are selected using invariant mass and flight distance criteria for
the intermediate hyperons, plus PID requirements on kaon and proton tracks.
The $p^*$ spectra shown in Figure~\ref{fig:Xic:pdist} are obtained by
subtracting background contributions, estimated using events from the mass
sidebands, and are not corrected for efficiency.
The peaks below 1.5~GeV/$c$ visible in the on-peak samples
(Figures \ref{fig:Xic:pdist}(c) and~(d)) provide clear confirmation of
$\Xi_c^0$ production in $B$ decays.

To measure the ratio of branching fractions to the two decay
modes, the candidates are further required to have $p^* > 1.8$~GeV/$c$.
After correcting for efficiency, the preliminary ratio of branching fractions is found to be
$0.296 \pm 0.018 \pm 0.030$, where the dominant systematic uncertainties are
from the fitting procedure and efficiency loss due to detector acceptance.
This improves upon the previous measurement by CLEO\cite{ref:Xic:cleo_paper}
and is consistent with a spectator quark model prediction\cite{ref:Xic:theory_prediction}.

\section{Conclusions}

\babar \ has a very active charm physics program. The topics discussed in
this paper are only a fraction of those presented at the
$32^{\rm nd}$ International Conference on High Energy Physics at Beijing,
and with the excellent luminosity achieved by PEP-II 
we expect many 
high-precision results to follow.

\end{document}